\documentclass[a4paper,12t]{article}
\usepackage[top=2.54cm,bottom=2.54cm, left=2.54cm, right=2.54cm]{geometry}
\usepackage{amsmath}
\usepackage{amsfonts}
\usepackage{amssymb}
\usepackage{graphicx}
\usepackage{bm}
\usepackage{enumerate}
\usepackage{color}
\usepackage[sort,comma]{natbib}
\usepackage{float}
\usepackage[colorlinks=true, allcolors=blue]{hyperref}
\usepackage{multirow}

\usepackage{tcolorbox}
\usepackage{xcolor,colortbl}

\definecolor{Gray}{gray}{0.85}
\definecolor{LightCyan}{rgb}{0.88,1,1}

\newcolumntype{a}{>{\columncolor{Gray}}c}
\newcolumntype{b}{>{\columncolor{white}}c}

\usepackage{alltt}

\title{Estimating NBA players salary share according to their performance on court: A machine learning approach}
\author{Ioanna Papadaki and Michail Tsagris\textsuperscript{*} \\
Department of Economics, University of Crete, Gallos Campus, Rethymnon, Greece  \\
\href{ipapadaki\textunderscore1998@yahoo.gr}{ipapadaki\textunderscore1998@yahoo.gr}
and \href{mailto:mtsagris@uoc.gr}{mtsagris@uoc.gr} \\
\textsuperscript{*} Corresponding author
}

\begin{document}

\maketitle

\begin{center}
{\bf Abstract}
\end{center}
It is customary for researchers and practitioners to fit linear models in order to predict NBA player's salary based on the players' performance on court. On the contrary, we focus on the players salary share (with regards to the team payroll) by first selecting the most important determinants or statistics (years of experience in the league, games played, etc.) and then utilise them to predict the player salaries by employing a non linear Random Forest machine learning algorithm. We externally evaluate our salary predictions, thus we avoid the phenomenon of over-fitting observed in most papers. Overall, using data from three distinct periods, 2017-2019 we identify the important factors that achieve very satisfactory salary predictions and we draw useful conclusions. 

\section{Introduction}
Professional athletes' field performance and salaries is a topic that has attracted the interest of numerous researchers \citep{zimmer2001,yilmaz2003,olbrecht2009,vincent2009,wiseman2010,garris2017}. The general question of interest is whether players deserve their salaries based on their performance statistics. We emphasize that this relationship is not linear and hence linear models are bound to fail in capturing the underlying true association. An additional concern, separate from non-linearity, is model predictability for which internal evaluation has limitations and leads to an over-optimistic performance. These and more matters, discussed later, require delicate treatment which, if not properly addressed, will yield erroneous results. 

Specifically for the NBA, \cite{sigler2000} studied the task of salary prediction using data from the 1997-1998 season but with only three predictor variables, rebounds, assists and points per game. \cite{ertug2013} related the players salaries with a set of predictor variables, most of which were not related to the players' performance on court. Their data were gathered from the 1989-1990 up to the the 2004-2005 period. More recently, \cite{xiong2017} performed a similar analysis using more predictor variables measuring the players performance on court for the 2013-2014 season. \cite{sigler2018} studied the 2017-2018 season but related the salaries with predictor variables exposing the players abilities on court. The ultimate question though is whether the results obtained from such analyses are valid and reliable enough. For instance, can a value of the coefficient of determination as high as 0.6 or 0.7 be a sign of correctness or even suggest that the analysis was successful?

To begin with, obtaining data is step zero, while pre-processing and "cleaning" them is the first step of the analysis. Standardization of the data is the first vital step, given that statistics are measured in different scales. For example, average points per game span between 0 and 30, whereas rebounds can reach values as high as 15-16 and average minutes per game vary between 0 and 40. However, caution must be taken on how and when to apply it. On a different direction, dealing with missing information and team changes within the same season are also part of the data cleaning process. 

The second step is to select the statistics or determinants with the highest effect on the player salaries. We advocate against the use of all available statistics and strongly encourage researchers to apply variable selection algorithms. Not only is it important to determine the appropriate statistics, but it is also crucial for the predictive performance of the models discussed below. We must further decide whether a single set of statistics (per game, per 36 minutes, etc.) or their combinations contain the highest amount (in linear terms) of information about the salaries and whether feature construction improves our predictions.

The third step is to apply sophisticated models to the selected statistics. Researchers ordinarily apply linear or generalised linear models (e.g. logistic regression), or cluster and discriminant analysis. The drawback of these models is their narrow abilities to capture non-linear components when the relationships among the variables are far from linear. Discriminant analysis with unequal group covariance matrices, is non-linear but is  specifically quadratic. This gives a higher degree flexibility than the discriminant analysis with equal group covariance matrices, yet is still not flexible enough for the associations of interest.

Adding these steps together leads to the two-fold aim of this paper. The first is the detection of the important statistics that incorporate the highest amount of information on the players' salaries. Do advanced statistics contain more information about the players than the per game or the per 36 minutes statistics? In either case, do we really require all sets of statistics or a subset of them? Selecting the appropriate statistics, not only removes the noise from the data, but also gives a better insight into the problem. Secondly, we predict the NBA player salaries using the selected statistics. What can we say about the relationship between the given statistics and the players' salaries? How accurate can our salary predictions be? We used the Least Absolute Shrinkage and Selection Operator (LASSO) to select the important statistics, while for the salary prediction we used the Random Forest (RF) algorithm. 

In the next section we describe the problem of NBA player salary prediction and provide information on the available statistics, how we pre-processed and "cleaned" the data and set the goal of this paper. In Section 4 we adumbrate improper approaches that are frequently followed. For instance, employment of linear models or erroneous application of the aforementioned non-linear algorithms to real and simulated data illustrate the occurrence of over-fitting, over-optimistic results. We describe the tools used for this purpose in the same Section; the variable selection we used to select the appropriate statistics and the machine learning algorithms we employed to predict the player salaries. We next delineate the proper approach, in Section 3, depicting how to properly evaluate the models by using the cross-validation (CV) procedure and present the results of our analysis. We further explain why our achieved predictive performance is the highest achieved ever by showing that all other methods have fallen in the pitfall of over-fitting. We finally summarise our findings concluding the paper.

\subsection{Description of the data} \label{description}
The starting point of the entire process is to compile all the required information about the players’ performance on court, their salaries, the team payrolls as well as other determinants that might prove useful. Our main source of data was \href{https://www.basketball-reference.com}{basketball-reference.com} which is broadly known for providing a great variety of reliable sports statistics. The data acquired were narrowed down to the NBA seasons of 2016-2017, 2017-2018 and 2018-2019. There were available statistics on 486 players for 2016-2017, on 540 for 2017-2018 and on 530 for 2018-2019.

Throughout the player statistics data accumulated, a multitude of 54 variables\footnote{There were some common variables though such as their age on February 1st of the season (Age), number of years they have played in NBA (EXP), the team (Tm) and the position (Pos) in which they played. additionally, the total number of games (G) and minutes (MP) they participated and the number of games they were in the starting five (GS) were displayed on almost all types of statistics.} provided a plurality of information about each players' performance per game, per 36 minuted and per 100 team possessions. Those include indexes for field goals, 3-point field goals and 2-point field goals counted as of total number (FG, 3P, 2P), total number of attempts (FGA, 3PA, 2PA) and percentage of successful attempts (FG\%, 3P\%, 2P\%). The index Effective Field Goal Percentage (eFG\%), found exclusively on the per game statistics, adjusts for the fact that a 3-point field goal is worth one more point than a 2-point field goal. In a similar manner, we have free throws (FT), free throw attempts (FTA) and free throw percentage (FT\%), offensive rebounds (ORB), defensive rebounds (DRB) and  total rebounds (TRB) per game, per 36 minutes and per 100 possessions. Furthermore, assists (AST), steals (STL), blocks (BLK), turnovers (TOV), personal fouls (PF) and points (PTS) were also included. Among the per 100 team possessions statistics two more variables were incorporated, Offensive Rating (ORtg) and Defensive Rating (DRtg), which are estimates of points produced by players or scored by teams per 100 possessions and of points allowed per 100 possessions respectively.

In the attempt to obtain a more comprehensive picture of the players' performances we also consulted the Advanced players' statistics available on \href{https://www.basketball-reference.com}{basketball-reference.com}. This type of statistics displays variables such as Player Efficiency Rating (PER), a measure of per-minute production standardized so that the league average is 15, True Shooting Percentage (TS\%), a measure of shooting efficiency that takes into account 2-point field goals, 3-point field goals, and free throws, 3-Point Attempt Rate (3PAr), which is the percentage of field goals attempts from a 3-point range and Free Throw Attempt Rate (FTr) which indicates the number of free throw attempts per field goal attempt. In addition, the percentage of available offensive rebounds, defensive rebounds and total rebounds a player grabbed while he was on the floor is estimated using Offensive Rebound Percentage (ORB\%), Defensive Rebound Percentage (DRB\%) and Total Rebound Percentage (TRB\%) respectively. 

Advanced statistics further comprise of Assist Percentage (AST\%), an estimate of the percentage of teammate field goals a player assisted, Steal Percentage (STL\%) and Block Percentage (BLK\%), estimates of the percentage of opponent possessions that end with a steal by the player and of opponent two-point field goal attempts blocked by the player, together with Usage Percentage (USG\%), an estimate of the percentage of team plays used by a player and Turnover Percentage (TOV\%), an estimate of turnovers committed per 100 plays. Moreover, we have estimates of the number of wins contributed by a player in total (Win Shares (WS)), due to his offense (Offensive Win Shares (OWS)), due to his defense (Defensive Win Shares (DWS)) and per 48 minutes (Win Shares Per 48 Minutes (WS/48)) with the last's league average being approximately 10\%. Lastly, we additionally incorporated the Offensive Box Plus/Minus (OBPM), Defensive Box Plus/Minus (DBPM) and Box Plus/Minus (BPM), which are box score estimates of the offensive, defensive and total points per 100 possessions a player contributed above a league-average player translated to an average team, along with the Value over Replacement Player (VORP), a box score estimate of the points per 100 team possessions that a player contributed above a replacement-level (-2.0) player, translated to an average team and prorated to an 82-game season.

Next on our data collection process, we turned to \href{http://www.espn.com/nba/salaries}{espn.com} and  \href{https://hoopshype.com/salaries/}{hoopshype.com} to attain the fundamental information about the players' income and the 30 NBA teams' payrolls for the seasons under investigation\footnote{ \href{https://hoopshype.com/salaries/}{Hoopshype.com} provided us with the choice between the absolute nominal value of each team's payroll or the payroll adjusted for inflation based on the current year (from data provided by the U.S. Department of Labor Bureau of Labor Statistic), to which we chose the first for the sake of correspondence between the base years on affiliated monetary values.}. As far as the players' salaries are concerned, for 2016-2017 the number of available observations was 594, for 2017-2018 598 and for 2018-2019 503. 

It is of major importance to take into account the teams' payroll when predicting the players' salaries, given the variation of this amount among different teams. During 2016-2017, Utah Jazz's payroll was the minimum across NBA with their contracts summing to \$80 millions, whereas Cleveland Cavaliers spent the highest amount of money, \$130 millions. During the 2017-2018 season Dallas had the minimum payroll with \$85 millions whereas Charlotte Hornets had the highest payroll of \$143 millions with Cleveland Cavaliers having second highest, \$137 millions. In our latest season, 2018-2019, Atlanta Hawks had the minimum payroll, whereas Miami Heat had the highest payroll, equal to \$79 millions and \$153 millions respectively. Markedly, had Miami Heat's best player signed with Atlanta Hawks he would earn around 65\% of his Miami salary, and conversely, if Atlanta Hawks' best player was traded to Miami he should get 150\% times his current salary. To make our predictions payroll free, instead of using the nominal wage for each player as the dependent variable on our regression models, we constructed a new variable, the ratio of the player's salary to his team's payroll. This is the players share salary. 

\cite{sigler2018} signified that a player's years on the league is a determinant of equal importance for his salary with his performance, as depicted by his statistics. He demonstrated the maximum amount of salary a player can receive based on the number of years he has played in the NBA and the amount of the salary cap. According to the 2017-2018 season, the maximum salary of a player with six or fewer years of experience is either \$25,500,000 or 25\% of the total salary cap, whichever is greater. For a player with 7–9 years of experience, the maximum increases to \$30,600,000 or 30\% of the salary cap, and for a player with 10+ years of experience, his maximum contract can reach \$35,700,000 or 35\% of the salary cap. There is also an exception to maximum salary in that a player can sign a contract for 105\% percent of his previous contract, even if the new contract is higher than the league limit. Having said this, we made use of \href{https://stats.nba.com/team/1610612738/}{stats.nba.com} to include each player's experience in our inquiry.

\subsection{Cleaning and pre-processing the data}
The volume of data accumulated needed to be merged into a unified database that would serve the purpose of our analysis. The objective of this process was to associate each player's wage with their statistics, their years of experience and their team's budget. However, we came across cases of missing information throughout the different sources. For example, there was no available salary or experience for some of the players listed on the statistics database and vise versa. To solve this problem, we solely kept the observations in which we had all three types of information at our disposal. Moreover, some of the players switched teams within the season and, as a result, they were recorded several times on the statistics database, once for every team. In this instance, we preserved the statistics exclusively for the team for which we had information about the salary. On account of better results, it was also deemed necessary to discard all players who participated in less than 10 games during each season, in view of the fact that those observations' contribution to our model's predictability was actually a drawback. Throughout this process of "cleaning" the data the remaining observations were 443 for 2016-2017, 484 for 2017-2018 and 412 for 2018-2019 which were considered to be adequate sample sizes.

As mentioned above, a new variable was constructed as the percentage of the player's salary to his team's payroll, which will later be used as the dependent variable on our models. The subtotal of this variable for each team ought to be $\leq 1$ and in cases this ratio exceeded 1 it was decided to replace the team's payroll with the aggregation of the team's players' salaries at hand. 

\subsection{Other possible determinants of salaries}
It can be argued that NBA player salaries are not only related to their performance on court but also to publicity and reputation \citep{ertug2013}. Reputation though is difficult to measure for all players, using for example followers in social networks, NBA contracts with sports companies, promoting activities, etc., and we thus have avoided it\footnote{For example, to measure a player's level of spectacle we would have to collect the number of dunks, the number of alley-hoops, the number of fake movements, the number of ankle-breaking phases or any other spectacular movements}. Other contributing factors include player managers that can make hard negotiations with team managers and can achieve higher earnings for their clients. We assert that these factors are projections of the players' image on the court. Highly skilled (and regularly spectacular) players are the ones that will ordinarily sign contracts with sport companies and will be interviewed and promoted by sports journalists.  

A second factor is discrimination, either racial \citep{kahn1988,hamilton1997,kahn2005,rehnstrom2009,wen2018} nationality-wise \citep{yang2012,hoffer2014} or exit \citep{groothuis2013}. Our personal view is that any alleged discrimination present is fully justifiable by the players' performance. African-American players have better physical skills and are more athletic, which facilitates the quantity of spectacle they offer compared to other players. If those players receive higher salaries simply because they may have better statistics or have a more spectacular type of play, this is by no means evidence of race or country discrimination. Further, foreign players, e.g. Europeans have nourished in a different mentality. American basketball is more athletic than European and usually Europeans require more adjustment time than players drafted from the NCAA. It is perceptible that athletic and physical abilities and the mentality of basketball has caused this alleged discrimination. Investigation of this entails a comparison of the player performance between African-American and white American players and between American and European players. The same rule applies for the exit discrimination and we quote a sentence from \cite{groothuis2013} \textit{From research on exit discrimination it is clear that individuals with greater ability have a higher survival rate}. We close the discrimination matter by referring to \cite{groothuis2013} who used a panel dataset from 1990 to 2008 and failed to find any evidence of either pay or exit discrimination in the NBA. 

A third possible determinant factor is TV contracts\footnote{TV contracts contribute to the NBA revenues which determines the salary cap.}, which were deemed as not important by \citep{kelly2017}. \cite{kelly2017} applied a linear regression model where a subset of the TV contracts relevant variables were statistically highly significant, yet the goodness of fit of the model was very poor\footnote{This is another case that exemplifies why non-linear models are necessary to yield more accurate predictions.}.

\section{Salary prediction}
We will now illustrate some incorrect approaches that are ordinarily followed by researchers. Subsequently we describe the pipeline for selecting the most important performance determinants that affect the player salaries and how to make the most of the predictive capabilities of those determinants. 

\subsection{Current approaches} \label{notgood}
\cite{sigler2000} related some player statistics (points, rebounds) with their salaries using a linear regression model and computed an unsurprisingly low coefficient of determination ($R^2$). \cite{ertug2013} used a linear regression model to estimate the team revenues attributed to ticket sales computing high $R^2$ values for two models, $0.75$ and $0.77$. When it came to estimating player salaries, their linear models had a low fit though ($R^2=0.30$ and $R^2=0.31$)\footnote{In hockey, \cite{vincent2009} applied a quantile regression instead, a robust to outliers regression model, but still linear.}. 
The fact that researchers do not select the important determinants prior to fitting the model is an example of what not to do. \cite{ertug2013} for example, in their seemingly optimal model for the team's ticket based revenue used 13 variables, out of which only 4 variables were statistically significant and 2 of them were highly statistically significant. Retaining the other 9 variables in the model does not add but removes value from it in a three-fold manner. a) It is known that the addition of variables in the model leads to higher $R^2$ values. Thus, the reported value of $0.77$ for the $R^2$ is an overestimate of the true $R^2$ of their model. b) This practice makes the model unnecessarily more complex and c) in fact deteriorates the predictive performance of the model. This is associated with the curse of dimensionality \citep{hastie2009} and is the main reason why variable selection is necessary, to remove the irrelevant variables that add noise and no information. Identically, national teams select their best among the all-star players when participate in international championships (Continental and Universal).

Attempting to model the non-linear relationships of the statistics with the player salaries via adoption of a linear model is a policy that should be avoided. A preferable strategy is to add of square terms in some variables may improve the performance of the model, but not significantly. Linear models will encapsulate, to some degree, the trend in the variables, but they definitely cannot be used for safe prediction. The following example from basketball suits to convey our message. NBA teams select the most talented rookie players, but solely talent is not enough. It is training that will take those players to the next level and the better the "material" in a teams hands the higher its chances to win the championship. 

Assessing the goodness of fit of a model via the $R^2$ is a criticism raising strategy. The performance of a model that has been constructed on some data must be tested on different data that the model has never "seen". During training, players test their abilities against one another, but soon they understand each other's play and perhaps some players perform very well, but only during practice. Players are not getting paid to play well during practice with their teammates, but to play well against new players whose team systems or play they have not seen. Players are always evaluated externally and not internally. 

Not all researchers though fall into the aforementioned pitfalls\footnote{Criticizing all available papers is outside the scope of this paper and hence we do not pursuit this further.}. \cite{wiseman2010} performed a variable selection procedure in order to predict the American League Baseball player salaries and also included a quadratic effect in the years of major league service. Howerver reporting an internal (and hence over-optimistic) $R^2$ as high as $70.1\%$ was an incorrect decision made by those researchers. 

\subsection{Variable selection and prediction algorithms}
To further assist the comprehension of the analysis we will narrate the LASSO variable selection and the RF algorithm\footnote{We performed the analysis using the open source software \textit{R} \citep{R2020}. LASSO is implemented in the R package \textit{glmnet} \citep{friedman2001}, while RF is implemented in \textit{ranger} \citep{ranger2017}.}.

\subsubsection{Least Absolute Shrinkage and Selection Operator}
Least Absolute Shrinkage and Selection Operator (LASSO) \citep{tibshirani1996} is a regression analysis method that performs variable selection and regularization of the regression coefficients. The objective of this method is to improve the prediction accuracy and interpretability of regression models by selecting a subset of the provided variables that exhibits the strongest effects on the response variable. LASSO is able to both improve prediction error by shrinking large regression coefficients in order to reduce overfitting, and perform variable selection, discarding variables that are responsible for large variance, therefore making the model more interpretable. 

LASSO minimizes the following penalised sum of squares
\begin{eqnarray} \label{lasso}
\sum_{i=1}^n\left(y_i - \sum_{j=1}^p\beta_jx_{ij}\right)^2 + \lambda\sum_{j=1}^p\left|\beta_j\right|,
\end{eqnarray}
where $y_i$ is the $i-th$ response value, $x_{ij}$ denotes the $i-th$ value of the $j-th$ predictor variable, $n$ denotes the sample size and $p$ is the number of predictor variables. Fine tuning the penalty parameter $\lambda$ is essential to the performance of LASSO since it determines the amount of regularization, the strength of shrinkage and, ultimately, the number of variables selected for use in the final model. Such is achieved through a cross-validation procedure, where the value $\lambda$ yielding the lowest estimated prediction error is preferred.

The correlations between NBA player salary shares and performance measures correlated statistically significantly, but not all of them remain significant when all predictor variables enter a regression model. The LASSO algorithm facilitates the detection of the most important performance measures.

\subsubsection{Random Forests}
The RF algorithm is a fast and flexible data mining approach well suited for high-dimensional data. The algorithm is built upon creating many classification or regression\footnote{Depending on the nature of the response variable.} trees. According to \cite{breiman2001} RF randomly draws a subset of variables and a bootstrap sample\footnote{Sample with replacement, of the same size.} and uses only this subset of features to grow a single tree. This process of randomly selecting variables and bootstrap samples is repeated multiple times and the results are aggregated. By creating many trees at random (500 or 1000 for instance), one ends up with a random forest. 

As stated in the Introduction, the relationship between the player salaries (shares) and their performance on court is not expected to be linear, hence the RF algorithm will allow us to capture the non-linear components of this relationship.

\subsection{A note on the response variable}
We stated earlier that we converted the player salaries into (payroll-free) percentages that are on the same basis for everyone. The implications of this transformation to LASSO, which employs a linear regression model, hence a normal distribution, are obvious. Unlike the normal distribution whose support is unbounded, the percentages lie within a restricted range of values. Additionally, predictions with LASSO are not constrained to lie within that plausible range. Not correct specification of a distribution or of a regression that takes into account the space where the response variable is defined is another frequent mistake of researchers and practitioners. 

We refrained from using the salary shares in LASSO and transformed the response prior to employing LASSO using the logit transformation $y^*=\log{\frac{y}{1-y}}$, where $y$ denotes the player salary shares. The logit transformation is well defined when $y\neq 0$ and $y \neq 1$ which holds true in our case. This transformation is not obligatory when RF are used since the predicted values are in fact weighted averages of the observed values \citep{lin2006}.

\subsection{Distribution of the NBA player salary shares}
\begin{figure}[!ht]
\centering
\includegraphics[scale = 0.47, trim = 20 0 0 0]{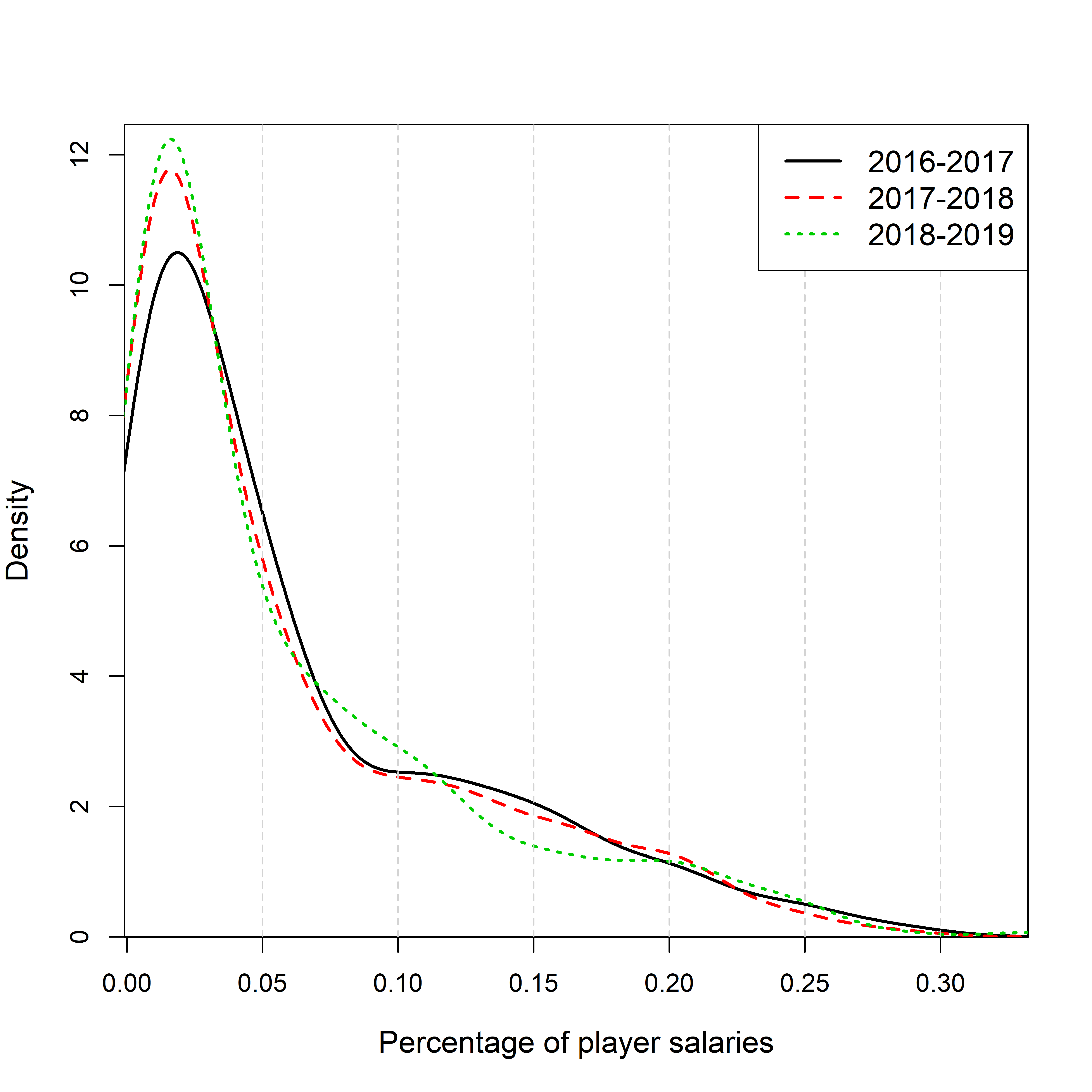} 
\caption{Kernel density estimates of the NBA player salary shares across the three seasons. \label{kde} }
\end{figure}

Figure \ref{kde} shows the kernel density estimates of the distributions of the salaries for each season. The differences are rather small and indeed there is no evidence to support that the distributions vary statistically significantly across the three seasons (p-value=0.9263). 

\subsection{Internal evaluation in our datasets}
We now illustrate the internal evaluation of RF in our datasets and manifest the over-rated performance they seem to be possessing. We standardised our predictor variables prior to the analysis in order to transform all variables into the same scale\footnote{We explained why this pre-processing step is incorrect in Section \ref{wrongstand}.}. We implemented the 10-fold CV procedure (described in the next section) to tune the penalty parameter ($\lambda$) of LASSO. We then performed LASSO penalisation using the chosen value of $\lambda$ to select the most important factors which we plugged in the RF algorithm. 

We evaluate the predictive performance of RF by contrasting each set of predictions (one set for each hyper-parameter) against the true salary shares using the Pearson correlation coefficient (PCC) (\ref{pcc}) and the percentage of variance explained (PVE) \footnote{PVE is equivalent to $R^2$ for the linear models.} (\ref{pve}).
\begin{eqnarray} \label{pcc}
PCC = \frac{\sum_{i=1}^n\left(y_i-\bar{y}\right)\left(\tilde{y}_i-\bar{\tilde{y}}\right)}{\sqrt{\sum_{i=1}^n\left(y_i-\bar{y}\right)^2}\sqrt{\sum_{i=1}^n\left(\tilde{y}_i-\bar{\tilde{y}}\right)^2}}
\end{eqnarray}
\begin{eqnarray} \label{pve}
PVE = 1 - \frac{\sum_{i=1}^n\left(y_i-\tilde{y}_i\right)^2}{\sum_{i=1}^n\left(y_i-\bar{y}\right)^2},
\end{eqnarray}
where $\tilde{y}_i$ refers to the predicted value of the $i$-th observation and $\bar{y}$ denotes the mean value. The PVE values for each dataset\footnote{We do not report number of variable splits that yielded the highest performance.} appear in Table \ref{overfit}. We also report the PVE values of LASSO as a comparison of the performance of a linear and of a non-linear algorithm. 

Unlike the mean squared error (MSE) or mean absolute error (MAE) the aforementioned metrics have a benchmark value to compare against. The raw values of MAE, or MSE do not reflect the performance of the model relative to model free predictions. On the contrary, for both PCC and PVE their maximum value is 1 indicating excellent predictive performance, whereas the minimal value equal of 0 refers to completely random predictions. Higher values of PCC indicate a higher number of correct model based predicted orderings, whereas higher values of PVE indicate that on average the errors of the model based predictions are much less than the errors of random, model free predictions.

\begin{table}[!ht]
\caption{PVE values for each dataset and each algorithm across the three seasons. \label{overfit}}
\begin{center}
\begin{tabular}{l|cc|cc|cc}  \hline \hline
\rowcolor{LightCyan!50}
Season                 &  \multicolumn{2}{c}{2016-2017} &  \multicolumn{2}{c}{2017-2018}  &
\multicolumn{2}{c}{2018-2019}  \\ \hline \hline
\rowcolor{LightCyan!50}
Dataset              & RF    & LASSO & RF  & LASSO & RF  & LASSO 
\\  \hline \hline
\rowcolor{lightgray!10}
Per game             & 0.910 & 0.420 & 0.869 & 0.246 & 0.884 & 0.220 \\ 
\rowcolor{lightgray!10}
Per 36 minutes       & 0.901 & 0.284 & 0.866 & 0.223 & 0.826 & 0.210 \\ 
\rowcolor{lightgray!10}
Per 100 possessions  & 0.900 & 0.305 & 0.866 & 0.223 & 0.826 & 0.164 \\ 
\rowcolor{lightgray!10}
Advanced data        & 0.848 & 0.163 & 0.842 & 0.139 & 0.862 & 0.181 \\  \hline \hline
\end{tabular}
\end{center}
\end{table}

We used the PVE\footnote{The predicted values of LASSO were first back-transformed to percentages using the inverse of the logit-transformation, $y=\frac{1}{1+e^{-y*}}$ and then we computed the PVE.} (\ref{pve}) for model assessment and perhaps the only safe conclusion we can draw from Table \ref{overfit} is that non-linear models have superseded the linear model of LASSO. RF always produced PVE values above 0.8 (or 80\%) indicating an excellent fit. If we compared these PVE values against the $R^2$ values reported in previous papers we would be delighted, not only because we outperformed their fit, but also because our PVE values are remarkably high. We will repeat ourselves that the cost of this high PVE is interpretability. RF do not produce a coefficient for each predictor variable that could reflect the variable's (marginal) effect on the salary shares. 

\subsection{Model complexity}
The last, but equally important, point to take into account is model complexity. Fitting a highly complex non-linear model does not necessarily yield better prediction. To demonstrate this we expose below a short script written in R\footnote{The example is reproducible and will always yield the same results.} evaluating internally its performance. We randomly generated a set of 20 predictor variables and a random response variable. We then applied a non-linear model (projection pursuit regression\footnote{We tried this model in our analysis but the results were not that accurate and hence we omitted them.}), where each time we increased the complexity of the model and computed the PCC (\ref{pcc}) between the observed and fitted values. 
\\
\begin{tcolorbox}[colback=white]
\begin{alltt}
set.seed(12345)
x <- matrix( rnorm(400 * 20), ncol = 20 )  ## randomly generated predictors
y <- rnorm(400)  ## randomly generated response
pcc <- numeric(10)
for (i in 1:10) \{
  mod <- ppr(y ~ x, nterms = i) ## Projection pursuit regression
  pcc[i] <- cor( fitted(mod), y)  ## Pearson correlation coefficient
\}
round(pcc, 3)
\textbf{0.443 0.575 0.644 0.666 0.748 0.736 0.844 0.933 0.860 0.950}
\end{alltt}
\end{tcolorbox}

Evidently, Pearson correlation between the observed and fitted values increases with model complexity. Further, surprisingly enough, we managed to obtain a high level of correlation when in fact there is no relationship between the response and the predictor variables. This again points out that an internal evaluation draws no safe conclusions as over-fitting occurs. A second source of complexity comes from the fact that we used all available 20 predictor variables and not a subset of them. This is an extra reason why we should have performed variable selection prior to estimating the predictive performance. Penalising for complexity, e.g. via Bayesian Information Criterion could have avoided this phenomenon, but even then, internal evaluation would over-estimate the true performance of this model. 

\section{A more valid approach}
The previous example indicates how we can get trapped in over-fitting. The reported PVE values refer to the internal evaluation of the models because these are internal PVE values. We will next elucidate the correct way to estimate a model's predictive performance (external evaluation) using the $k$-fold CV protocol. To obtain an unbiased estimate of a model's predictive performance we need large sample sizes, a condition we meet because we have information on hundreds of players at each season. Finally, we will demonstrate that the observed performance metrics in Table \ref{overfit} are actually very high and far from reality.

\subsection{The $k$-fold CV} \label{kfoldcv}
The $k$-fold CV protocol is based upon splitting the data into $k$ mutually exclusive groups, termed folds. The ordinary value of $k$, which we also used, is $k=10$, yielding the $10$-fold CV. We select one fold and leave it aside to play the role of the test set. The remaining $9$ folds are combined into what is called the training set. We standardize the predictor variables of the training set only. We then perform variable selection using LASSO and feed the RF algorithm with the selected variables. We use the same selected predictor variables from the test set and we scale them using the means and standard deviations of the same predictor variables from the training set. We use these scaled predictor variables of the test set to predict the values of the response variable (percentages of player salaries) of the test set. For RF we used a range of splits of variables\footnote{These are termed hyper-parameters and need to be tuned.}. Thus, we end up with multiple predictions, one set of predictions for each hyper-parameter, whose predictive performance we compute.

We subsequently select another fold to play the role of the test set and insert the previous fold - (previous test set) into the training set and repeat the pre-described pipeline. The process is repeated until all folds have played the role of the test set. In the end, we collect all predictions from each fold resulting in an $n \times M$ matrix, where $n$ is the sample size and $M$ is the number of hyper-parameters, the total number of splits of variables, corresponding to $M$ sets of predictions. We compute the average predictive performance of each hyper-parameter separately and choose the hyper-parameter with the highest predictive performance. 

\subsection{The essence of CV}
The importance and necessity of any CV protocol can be further appreciated through an investment example. Assume an NBA team manager or team owner who wishes to invest their money on some market, stock exchange, mutual or pension funds, real estate, etc. There are two available investment companies residing in the building right next to his/her. Company A has a long record of remarkably high PVE values in their models. The company gathers the prices spanning from several days ago up to today and fits variable selection and machine learning algorithms and computes the PVE values of the models/algorithms using the same data. It shows no record of predicting future prices though. Company B on the other hand applies a different strategy. It again uses historical data, but keeps the old ones for model building and training and treats the most recent ones as the future that must be predicted. The PVE values (of the future predictions) of company B are significantly lower than the PVE values (of the past and present predictions) of company A but are safer predictions of the future. Company A implements the wrong approach described in Section \ref{notgood}, whereas company B implements the correct strategy described this section.  

\subsection{The importance of processing the data in the training set} \label{wrongstand}
CV can be seen as a simulation of realistic scenarios. Let us denote the training set by \textit{present} and the test set by \textit{future}. We observe the present and attempt the predict the future. We process (standardise) the data in the training set (present) and use those means and standard deviations to scale the test data (future). Had we standardised the data from the beginning would deviate from the realistic scenario as we would have allowed information from the future to flow into the present. Thus, attempting to transform the data into the same scale prior to performing any CV protocol is erroneous and should be avoided. 

But why is standardization so important? Numerous variables listed on the per game, per 36 minutes and per 100 possessions refer to percentages therefore deviate between 0 and 1, games played (G) can reach values as high as 82 and players' ages vary between 19 and 42. Furthermore, 3-point field goals per 100 team possessions (3P), for example, span between 0 and 7,2 and total rebounds per 100 team possessions (TRB) between 3,0 and 23,8. Likewise, the majority of the advanced statistics are estimates of percentages, while Win Shares (WS) are measured on a scale of -1,7 to 15,4 and Box Plus/Minus (BMP) of -5,7 to 11,1 ,just to name a few. Standardization is a necessary processing strategy in order to prevent our results from being strongly affected by the scale of measurement of the variables\footnote{This is the reason why VS algorithms, such as LASSO, require standardised data.}. 

\subsection{Results of the $10$-fold CV protocol}
Unarguably partitioning the data into 10 folds contains an inherent variability as different partitions will give different results. To robustify our inference against this uncertainty we repeated the $10$-fold CV procedure (variable selection and predictive performance estimation) 50 times and report the aggregated predictive performance of the RF algorithm. 

\begin{figure}[!ht]
\centering
\begin{tabular}{cc}
\includegraphics[scale = 0.47, trim = 50 0 0 0]{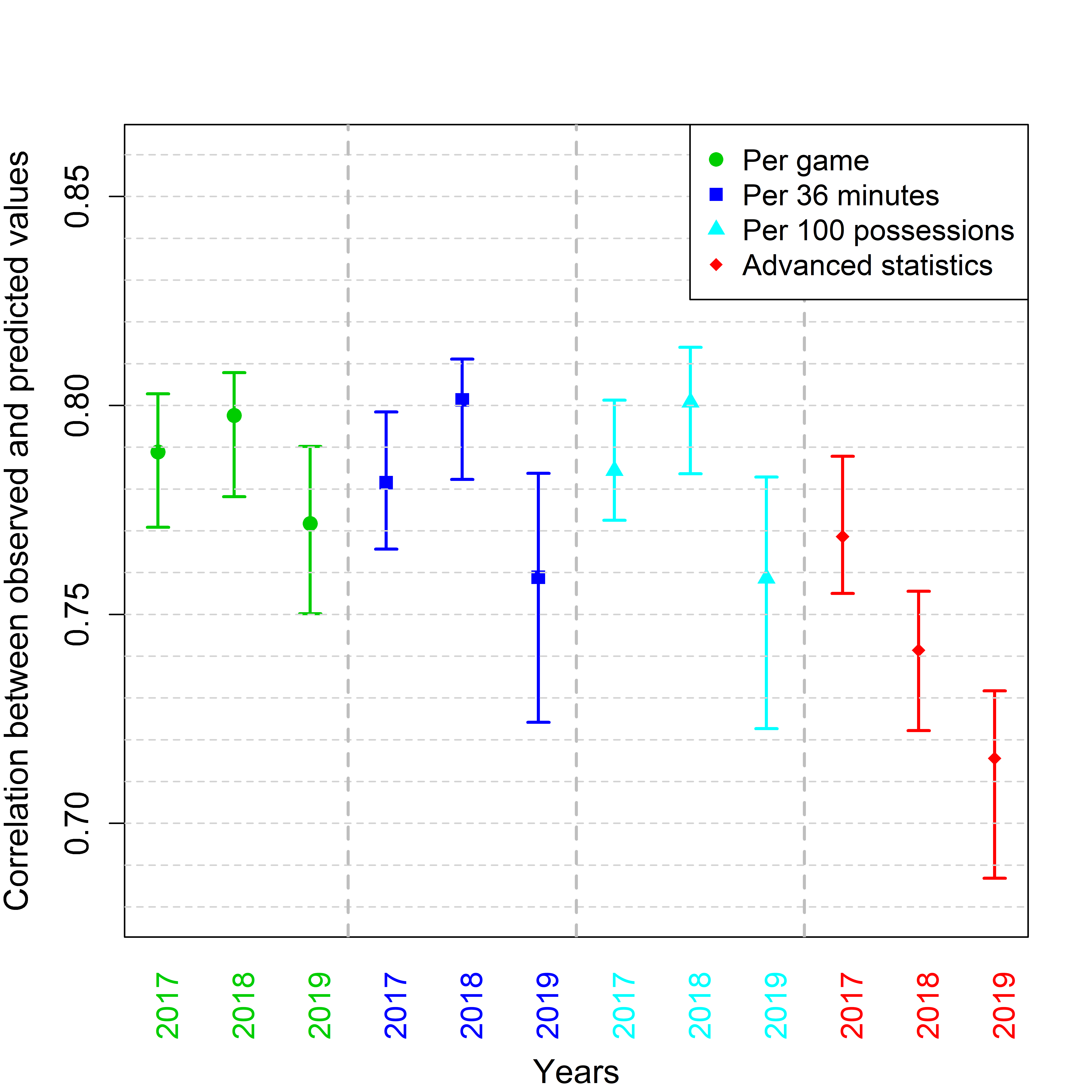}  &
\includegraphics[scale = 0.47, trim = 20 0 0 0]{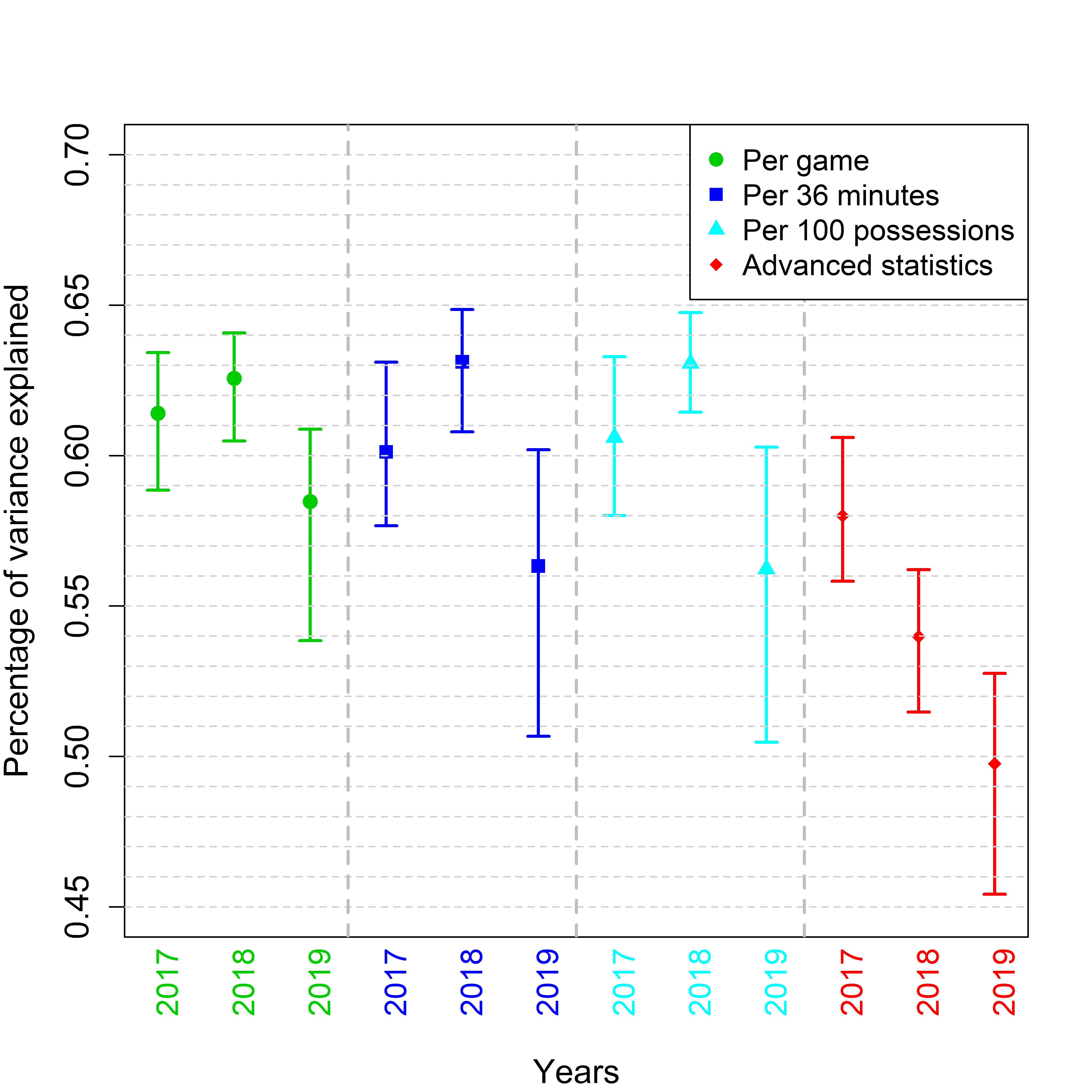}  \\
(a) PCC (\ref{pcc})  &   (b)  PVE (\ref{pve})
\end{tabular}
\caption{Predictive performance metrics using each dataset across the three seasons. \label{fig} }
\end{figure}

Figure \ref{fig} contains the average PCC (\ref{pcc}) and PVE (\ref{pve}) for every season using either set of statistics (Per game, Per 36 minutes, Per 100 possessions and Advanced statistics), along with the corresponding 95\% confidence intervals. Overall, use of the advanced statistics resulted in the worst performance among all datasets while the \textit{Per 36 minutes} and \textit{Per 100 possessions} portrayed a very similar picture, perhaps due to the fact that LASSO was selecting the same statistics. The \textit{Per game} statistics evidently gave the optimal predictions overall with an exception for the season 2017-2018, whose predictions ranked second best with the difference being tiny. The \textit{Per game} statistics also dominated in terms of variance of the predictive performance. The length of the 95\% confidence intervals for the true predictive performances are always the shortest, indicating higher stability.

There is a common pattern among the first three sets of statistics. We observe an increase in the predictability as we move from the 2016-2017 to the 2017-2018 season which then decays as we move to the 2018-2019 season. Further, it is in that last season that we observe the highest variability in the predictive performance and the confidence intervals are the widest observed across the three seasons. 

Tables \ref{tabpcc} and \ref{tabpve} present the optimal (average) predictive performances of the RF using each dataset across the three seasonss when LASSO variable selection has been applied prior to RF and when all statistics were fed into the RF. The PCC values are remarkably high, lying in the range of $0.7-0.8$. The PVE values on the other hand naturally take lower values, yet these number are high in comparison to prior research and most importantly those numbers were produced by external and not internal evaluation. Further, these two tables clearly visualise the essence of variable selection prior to using RF. The predictive performance changes slightly but LASSO selects the most important statistics, presented in Table \ref{stab}.

Table \ref{stab} contains the statistics that were most frequently selected by LASSO throughout the 50 repetitions of the 10-fold CV. Overall, experience and minutes played of each player were the two statistics that were always selected regardless of the set of statistics and the year of play. The third statistic was either the games played or the games started, followed by the points, the defensive rebounds and the field goals attempted. In the \text{Advanced statistics} the USG (an estimate of the percentage of team plays used by a player) and OBPM (box score estimate of the offensive, defensive and total points per 100 possessions a player contributed above a league-average player translated to an average team). Excluding the \textit{Advanced statistics}, we can see that there seems to be a stability in the selected statistics across the three seasons. In the \textit{Per game}, the last season only substitutes the games played, the filed goal attempts and the defensive rebounds with the points scored. A common feature with the \textit{Per 36 minutes} is that defensive rebounds do not seem to play a significant role in the last season. When it comes to the \textit{Per 100 possessions}, defensive rebounds never seem to contribute to the salary of the players. 

The use of advanced statistics did not yield better results than the use of the per game statistics, in fact the former dataset produced the worst results. We emphasise that the PER index is included in the advanced statistics.

\begin{table}[!ht]
\caption{PCC values for each dataset across the three seasons. \label{tabpcc}}
\begin{center}
\begin{tabular}{l|ccc|ccc}  \hline \hline
\rowcolor{LightCyan!50}
          & \multicolumn{3}{c}{With LASSO}  &  \multicolumn{3}{c}{Without LASSO} \\ \hline
\rowcolor{LightCyan!50}
Dataset   & 2016-2017  &  2017-2018  &  2018-2019  &  2016-2017  &  2017-2018  &  2018-2019  \\ \hline \hline
\rowcolor{lightgray!10}
Per game             &  0.789  &  0.798  &  0.772  &  0.783  &  0.794  &  0.758  \\    
\rowcolor{lightgray!10}
Per 36 minutes       &  0.781  &  0.801  &  0.759  &  0.772  &  0.781  &  0.750  \\
\rowcolor{lightgray!10}
Per 100 possessions  &  0.784  &  0.801  &  0.759  &  0.769  &  0.778  &  0.747  \\
\rowcolor{lightgray!10}
Advanced statistics  &  0.769  &  0.741  &  0.716  &  0.742  &  0.752  &  0.718  \\
\hline \hline
\end{tabular}
\end{center}
\end{table}

\begin{table}[!ht]
\caption{PVE values for each dataset across the three seasons. \label{tabpve}}
\begin{center}
\begin{tabular}{l|ccc|ccc}  \hline \hline
\rowcolor{LightCyan!50}
          & \multicolumn{3}{c}{With LASSO}  &  \multicolumn{3}{c}{Without LASSO} \\ \hline
\rowcolor{LightCyan!50}
Dataset   & 2016-2017  &  2017-2018  &  2018-2019  &  2016-2017  &  2017-2018  &  2018-2019  \\ \hline \hline
\rowcolor{lightgray!10}
Per game             &  0.614  &  0.626  &  0.585  &  0.600  &  0.616  &  0.561  \\        
\rowcolor{lightgray!10}
Per 36 minutes       &  0.601  &  0.631  &  0.563  &  0.577  &  0.589  &  0.540  \\
\rowcolor{lightgray!10}
Per 100 possessions  &  0.606  &  0.631  &  0.562  &  0.572  &  0.584  &  0.534  \\
\rowcolor{lightgray!10}
Advanced statistics  &  0.580  &  0.540  &  0.498  &  0.534  &  0.548  &  0.499  \\
\hline \hline
\end{tabular}
\end{center}
\end{table}

\begin{table}[!ht]
\caption{Most important statistics per dataset across the three seasons.  \label{stab}}
\begin{center}
\begin{tabular}{l|lll}  \hline \hline
\rowcolor{LightCyan!50}
Dataset   & 2016-2017  &  2017-2018  &  2018-2019   \\ \hline \hline
\rowcolor{lightgray!10}
Per game             &  EXP, MP, G, FGA, DRB  &  EXP, MP, G, FGA, DRB  &  EXP, MP, PTS   \\   
\rowcolor{lightgray!10}
Per 36 minutes       &  EXP, MP, GS, DRB, PTS  &  EXP, MP, GS, DRB, PTS  &  EXP, MP, GS, PTS  \\
\rowcolor{lightgray!10}
Per 100 possessions  &  EXP, MP, GS, PTS       &  EXP, MP, GS, PTS       &  EXP, MP, GS, PTS   \\
\rowcolor{lightgray!10}
Advanced statistics  &  EXP, MP, USG., OBPM    &  EXP, MP,  USG., OBPM   &  EXP, MP, USG., OBPM  \\
\hline \hline
\end{tabular}
\end{center}
\end{table}

\subsection{Testing the predictability of the RF algorithm}
\begin{figure}[!ht]we 
\centering
\begin{tabular}{cc}
\includegraphics[scale = 0.47, trim = 50 0 0 0]{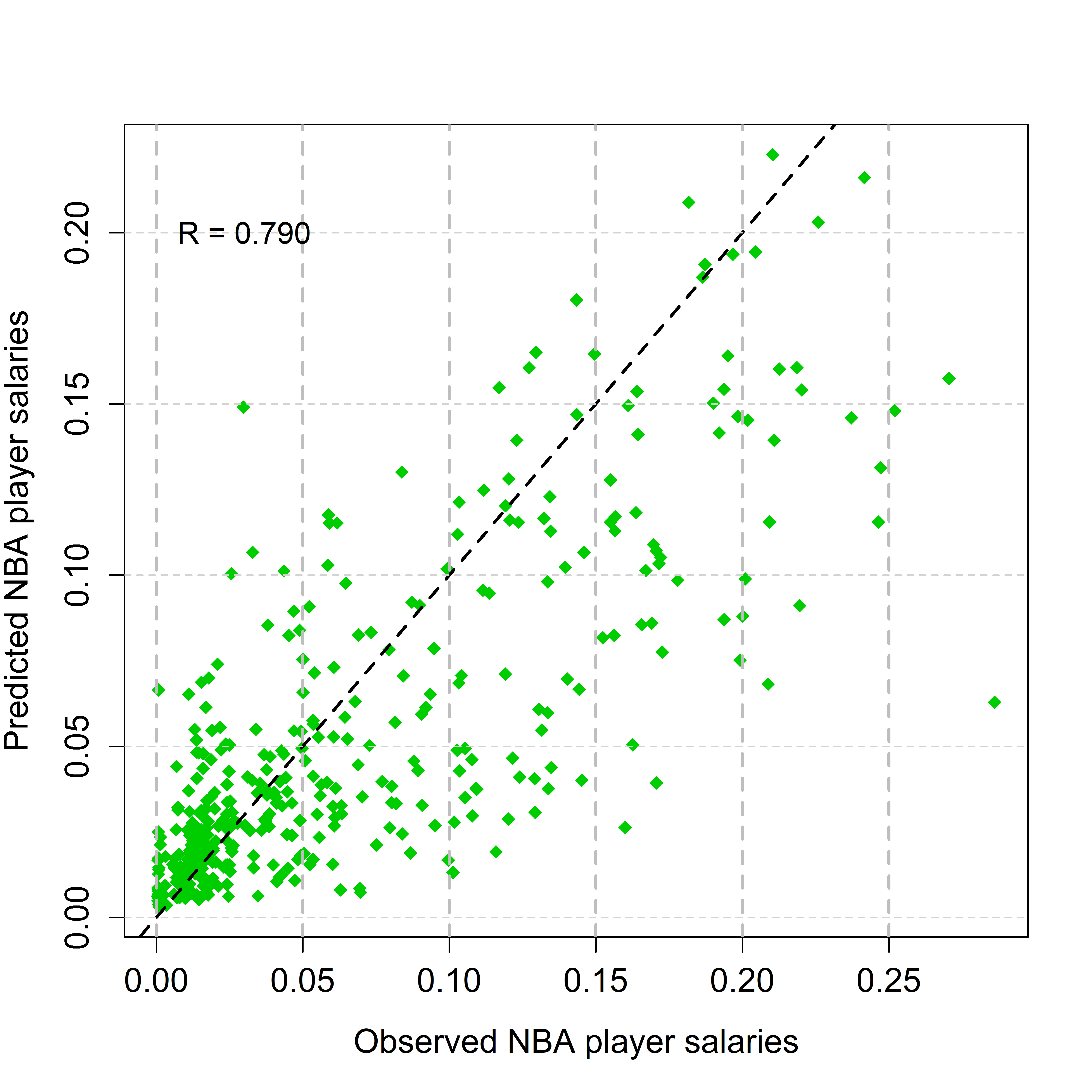}  &
\includegraphics[scale = 0.47, trim = 20 0 0 0]{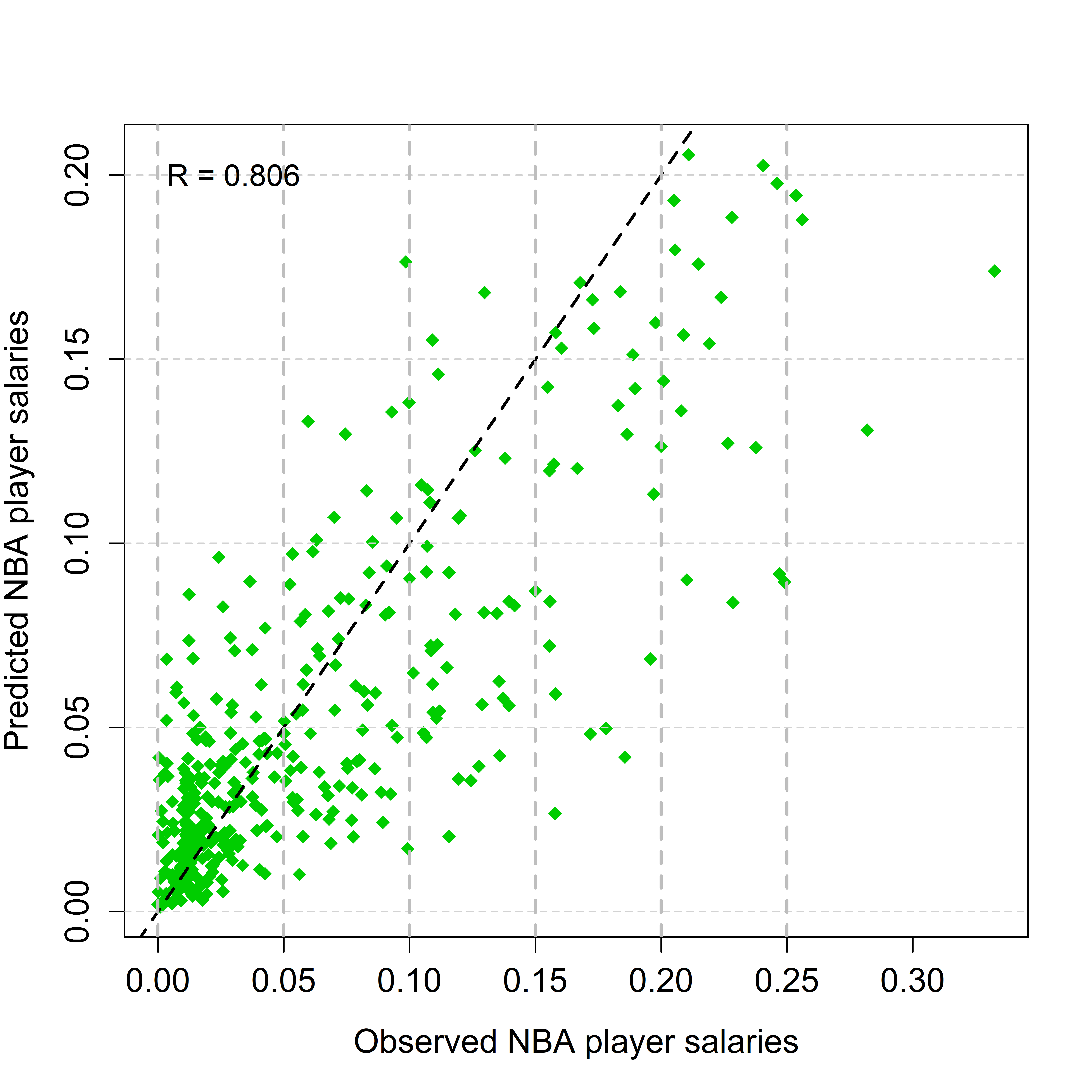}  \\
(a)  2017-2018  &   (b) 2018-2019
\end{tabular}
\caption{Observed vs predicted player salary shares for 2017-2018 and 2018-2019. \label{preds} }
\end{figure}

In order to show the validity of the PCC and PVE values reported in Tables \ref{tabpcc} and \ref{tabpve} respectively, we used the identified statistics from the 2016-2017 season, fitted an RF and predicted the salary shares of the 2017-2017 season. We repeated the same task using the statistics from the 2017-2018 season to predict the salary shares of 2018-2019. This way the next season's data played the role of the validation set, a new dataset that was never "seen" by the algorithms during the CV protocol. The PVE values were 0.624 and 0.650 respectively, while the PCC values were equal to 0.790 and 0.806 respectively. The observed and the predicted salary shares are displayed in Figure \ref{preds}. 

\subsection{NBA player salary share classes}
Let us now provide some in depth statistics regarding the player salaries. Table \ref{percent} shows the distribution of the player salaries across the three seasons.  

\begin{table}[!ht]
\caption{Distribution of player salaries across the three seasons. \label{percent}}
\begin{center}
\begin{tabular}{l|cccccc}  \hline \hline
\rowcolor{LightCyan!50}
          & \multicolumn{6}{c}{Player salaries in percent of the teams payroll} \\
\rowcolor{LightCyan!50}
Season    &  [0, 5\%)  & [5\%, 10\%)  & [10\%, 15\%)  & [15\%, 20\%)  & [20\%, 25\%)  & [25\%, 30\%]   \\ \hline \hline
\rowcolor{lightgray!10}
2016-2017   &  239  &  73  &  48  &  32  &  17  &  6  \\
\rowcolor{lightgray!10}
2017-2018   &  257  &  66  &  49  &  33  &  17  &  3  \\
\rowcolor{lightgray!10}
2018-2019   &  231  &  75  &  39  &  25  &  18  &  4  \\  \hline \hline
\rowcolor{lightgray!10}
\end{tabular}
\end{center}
\end{table}

During the 2016-2017 season, the 6 highest paid players (in terms of team's payroll share) were James Harden (Houston Rockets point guard, 29.18\%), Al Horford (Boston Celtics center, 28.40\%), Russell Westbrook (Oklahoma City Thunder point guard, 26.75\%), Kevin Durant (Golden State Warriors power forward, 26.13\%), Brook Lopez (Brooklyn Nets center, 25.69\%) and Dwyane Wade (Chicago Bulls shooting guard, 25.08\%). Amog them, James Harden was second in the points per game (29.1), Russell Westbrook and Kevin Durant's statistics justify their salaries. Surprisingly enough, Al Horford's statistics do not match his salary, as he was scoring 14 points per game despite playing 32 minutes. The same is true for brook Lopez and Dwyane Wade whose statistics are rather low.  

During the 2017-2018 only 3 players received more than 25\% of the team's payroll, Paul Millsap (Denver Nuggets power forward, 28.61\%), Harrison Barnes (Dallas Mavericks power forward, 27.05\%), and Stephen Curry (Golden State Warriors point guard, 25.20\%). Stephen Curry was scoring an average of 26.4 points per game, whereas Paul Millsap and Harrison Barnes were as low as 14.6 and 18.9 points per game, despite playing 30 and 34 minutes per game respectively.  

The 4 highest paid players (in terms of salary shares) for the last season, 2018-2019, were Lebron James (Los Angeles Lakers small forward, 33.25\%), Chris Paul (Houston Rockets point guard, 28.19\%), Stephen Curry Golden State Warriors point guard, 25.60\%) and Blake Griffin (Detroit Pistons, power forward, 25.36\%). Chris Paul was the only one among those 4 to score less than 20 points per game (15.6) even though he was playing 32 minutes per game. He was giving 8.2 assists per game and stealing the ball 2 times per game, yet, these statistics do not match that large salary share. 

The aforementioned players were evidently receiving a remarkably high share of their team's payroll, more than a quarter. Lebron James received an excessively high share, more than a third of Los Angeles Lakers' payroll, during the 2018-2019 season. We have no evidence to conclude that the highest paid players belong to the champion team. Kevin Durant won the NBA championship with the Golden State Warrriors in 2017 and Stephen Curry was a member of the same team that won the championship in 2018. Toronto Raptors won the championship, but their best player\footnote{Kawhi Leonard was receiving 16.78\% of Toronto Raptors' payroll.}, is not in the aforementioned list. These players are not the best among NBA and this small piece of information markedly shows that salaries are not always affected by statistics, hence partially explaining why salary prediction is hard to do with only performance statistics.

\begin{figure}[!ht]
\centering
\begin{tabular}{cc}
\includegraphics[scale = 0.47, trim = 50 0 0 0]{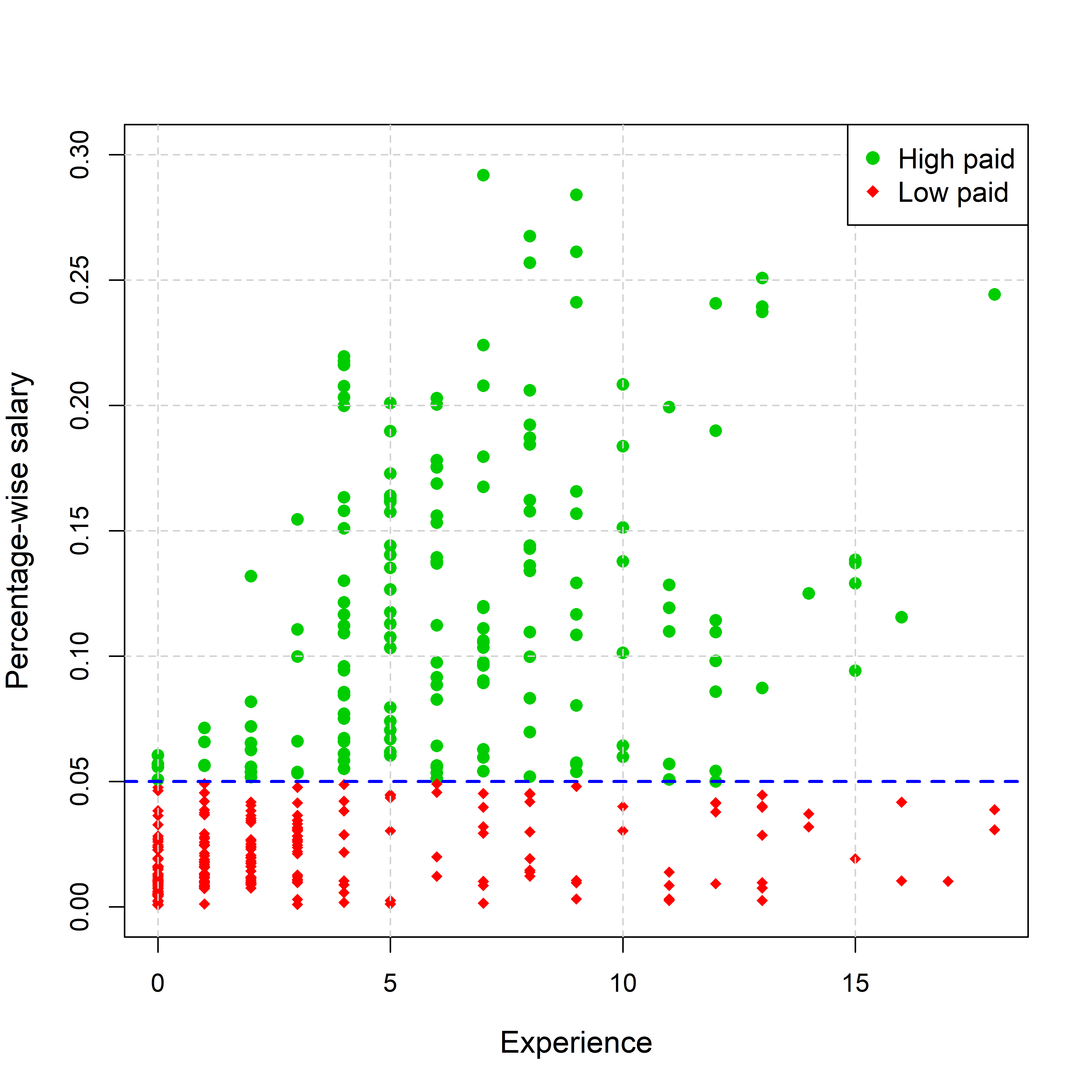}  &
\includegraphics[scale = 0.47, trim = 20 0 0 0]{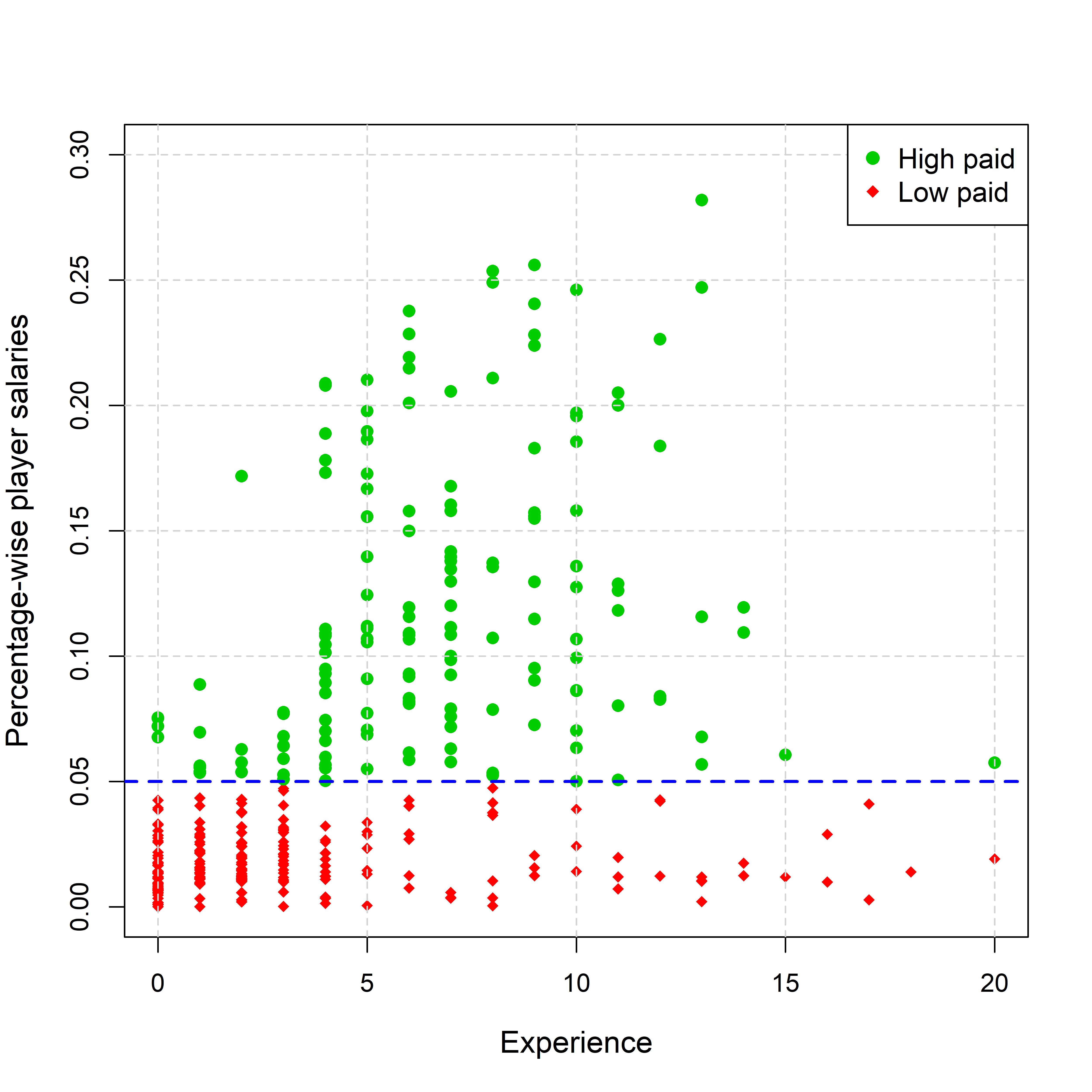}  \\
(a) 2016-2017 season  &   (b)  2018-2019 season
\end{tabular}
\caption{Player salary shares against number of years in the NBA. \label{fig2} }
\end{figure}

Having mentioned earlier that the number of years in the NBA affects the player salaries we visualize their relationship in Figure \ref{fig2}\footnote{We present this information for the 2016-2017 and 2018-2019 seasons only, due to space limitations. The scatter plot for the 2017-2018 season, and the scatter plots for the number of games played and the number of games the players were in the starting five were similar and hence omitted.}. Their relationship is clearly non-linear, the Pearson correlations are rather low (0.45 and 0.42 respectively) and there is no apparent threshold to separate the low from the highly plaid players. We cannot visually distinguish, in a straightforward manner, the low from the highly paid players. Further, broadly speaking, there is tendency for the salaries to increase, as expected, with thee years of service in the league but, percentage-wise this is not true for all players. 

\subsubsection{Salary share class prediction}
Table \ref{percent} transparently presents that most NBA players receive a small percentage (at most 5\%) of the teams payroll. This led us to the second part of our analysis, that of discriminating between the low and the higher paid players. To this end we employed the LASSO and RF algorithms again. In this scenario LASSO selects the most appropriate statistics by minimizing a more appropriate penalised function
\begin{eqnarray} \label{logi}
\sum_{i=1}^n\left[y_i \sum_{j=1}^p\beta_jx_{ij} - \log{\left(1+e^{\sum_{j=1}^p\beta_jx_{ij}} \right)} \right] + \lambda \sum_{j=1}^p|\beta_j|,
\end{eqnarray}
where $y$ takes two values, 0 and 1 corresponding to players receiving at lower than 5\% of their team's payroll or more, respectively. 

\subsubsection{Assessment of the classification task}
We used the Area Under the Curve (AUC) as the performance metric in this scenario. AUC represents the probability of correctly classifying a sample to the class it belongs to, thus takes values between 0 and 1, where 0.5 denotes random assignment. Unlike the accuracy metric (proportion of correctly classified samples), AUC is not affected by the distribution of the two classes (low or highly paid players). 

\begin{figure}[!ht]
\centering
\includegraphics[scale = 0.47, trim = 50 0 0 0]{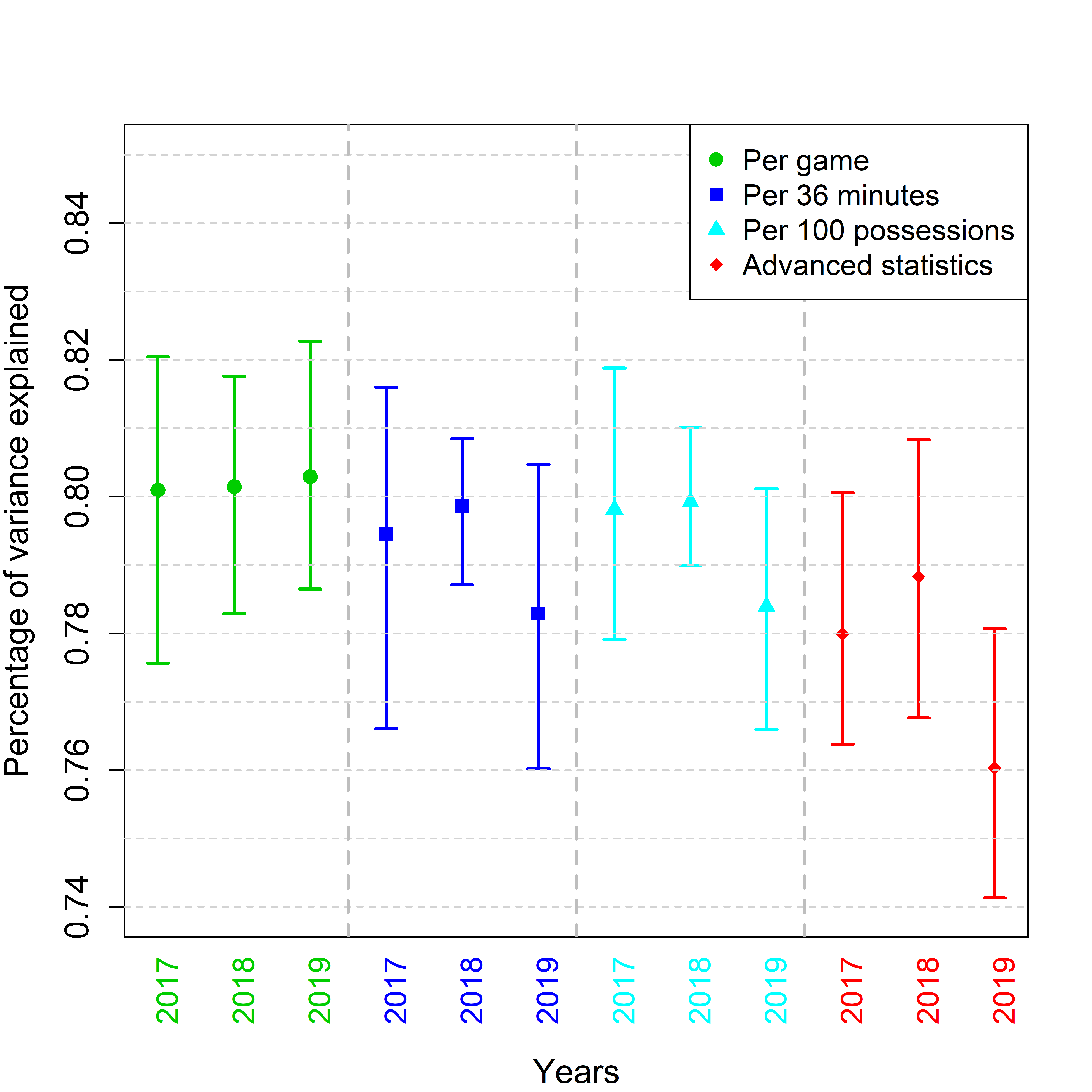} 
\caption{AUC using each dataset across the three seasons. \label{figauc} }
\end{figure}

We implemented the same repeated (50 times) $10$-fold CV protocol and present the results  in Figure \ref{figauc}. Once again, the \textit{Per game} statistics resulted in the optimal predictive performance, for which the average AUC was always greater than 0.80, whereas the \textit{Advanced statistics} yielded the lowest predictive performance. To appreciate the significance of this high value we can give the following interpretation. Knowledge of the number of years a player has played in the league and the average number of minutes he played for a given season allows to classify him his group (low or highly paid) with a probability equal to 0.8.

In terms or the selected statistics, LASSO was consistently selecting the same statistics as can be seen in Table \ref{stab2}. The number of years the player has played in the league, the average minutes they played in each game and the number of games they were in the starting five were the most important statistics throughout the datasets and the three seasons. 

\begin{table}[!ht]
\caption{Most important statistics per dataset across the three seasons.  \label{stab2}}
\begin{center}
\begin{tabular}{l|lll}  \hline \hline
\rowcolor{LightCyan!50}
Dataset              & 2016-2017       &  2017-2018    &  2018-2019   \\ \hline \hline
\rowcolor{lightgray!10}
Per game             &  EXP, MP        &  EXP, MP, GS  &  EXP, MP      \\   
\rowcolor{lightgray!10}
Per 36 minutes       &  EXP, MP, GS    &  EXP, GS      &  EXP, MP, GS   \\
\rowcolor{lightgray!10}
Per 100 possessions   &  EXP, MP, GS   &  EXP, GS      &  EXP, MP, GS   \\
\rowcolor{lightgray!10}
Advanced statistics   & EXP, MP, OBPM  &  EXP, MP, WS  &  EXP, MP, WS   \\
\hline \hline
\end{tabular}
\end{center}
\end{table}

As a second, validation, step we used the selected statistics from the 2016-2017 season, namely the number of years in the league and the minutes played, fed them into an RF using the statistics of 2017-2018 and predicted the salary share class of that season. We repeated this task to predict the salary share classes of the 2018-2019 season. The reasoning behind is to test the algorithm's ability to predict the next season's salary share classes. The AUC values for the 2017-2018 and the 2018-2019 predictions were 0.811 and 0.841 respectively, corroborating the results of the CV process. 

\subsection{Further analysis}
We further performed other variable selection algorithms (gOMP, \cite{tsagris2020}) and non-linear prediction algorithms such as projection pursuit \citep{friedman1981} and $k-NN$ \citep{altman1992} but their results were sub-optimal and hence omitted. Additionally we performed variable construction and combination of all datasets. For each dataset we constructed more variables, such as square and cubic transformation of each variable, along with all pairwise products of the variables. The second strategy was to combine all variables. The third strategy was to use all variables for each dataset and ignore the variable selection phase. None of these strategies improved the predictive performance of the RF.

\section{Conclusions}
The relationship between NBA player statistics and their salaries (expressed as percentage of the team's payroll) is evidently non-linear and we showed the necessity to apply non-linear models and algorithms. Using real and simulated data we showed the erroneous decisions made when based on applying linear models and the over-optimistic results, that even non-linear models yield when they are internally validated. We also showed the correct approach to investigate the relationship between a response and many predictor variables and how to estimate a model's predictability. 

Using the LASSO variable selection we managed to detect the important factors (statistics) that are mostly associated with the NBA player salaries and using the RF non-linear algorithm we predicted the player salaries satisfactorily enough. The level of achieved accuracy is the highest observed ever, to the best of our knowledge. The validity of the variable selection process and non-linear prediction was evaluated using a repeated cross-validation protocol yielding reliable results. 

Predicting NBA player salaries using information on the players' performance on court yields predictions whose accuracy is satisfactory but not as high as one would expect. We argue that key factors mentioned in the manuscript, such as popularity, quantity of spectacle offered, etc. could improve the accuracy of the salary predictions significantly. Another future idea is to switch direction. Instead of investigating the present, whether the players are getting paid according to what they present on court, one should investigate their future salaries. The level of the contract of a free agent depends upon his record but also upon many factors, his age, his playing position and the available teams among others. For instance, a center with high performance will sign with a team that is looking for a center. Additionally, among those teams interested in that player, one must see their salary cap and the players already in that team in order obtain a better picture. Further, we did not include more personal information, such as whether a player is an All-Star, if he is a member of the all-NBA team or the NBA All-Defensive Team, etc. Examination of all those factors could yield more accurate salary predictions than those presented in this paper. Adoption of more complex machine learning algorithms, such as SVM \citep{drucker1997} or gradient boosting \citep{friedman2001} is another possibility worth exploring.

We close this paper by posing a question. Is it possible that more than one combinations of statistics facilitate the prediction of the NBA player salaries? Evidently, the minutes played, the field goals attempted and the points scored are correlated. By observing the selected statistics in Table \ref{stab} we saw that the points scored, substituted the games played and the field goals attempted, only for the last season. This could be evidence that the variable selection task returns one solution among the many.


\begin{thebibliography}{}

\bibitem[\protect\citeauthoryear{Altman}{Altman}{1992}]{altman1992}
Altman, N.~S. (1992).
\newblock {An Introduction to Kernel and Nearest-Neighbor Nonparametric
  Regression}.
\newblock {\em The American Statistician\/}~{\em 46\/}(3), 175--185.

\bibitem[\protect\citeauthoryear{Breiman}{Breiman}{2001}]{breiman2001}
Breiman, L. (2001).
\newblock Random forests.
\newblock {\em Machine Learning\/}~{\em 45\/}(1), 5--32.

\bibitem[\protect\citeauthoryear{Drucker, Burges, Kaufman, Smola, and
  Vapnik}{Drucker et~al.}{1997}]{drucker1997}
Drucker, H., C.~J. Burges, L.~Kaufman, A.~J. Smola, and V.~Vapnik (1997).
\newblock Support vector regression machines.
\newblock In {\em Advances in Neural Information Processing Systems}, pp.\
  155--161.

\bibitem[\protect\citeauthoryear{Ertug and Castellucci}{Ertug and
  Castellucci}{2013}]{ertug2013}
Ertug, G. and F.~Castellucci (2013).
\newblock {Getting what you need: How reputation and status affect team
  performance, hiring, and salaries in the NBA}.
\newblock {\em Academy of Management Journal\/}~{\em 56\/}(2), 407--431.

\bibitem[\protect\citeauthoryear{Friedman}{Friedman}{2001}]{friedman2001}
Friedman, J.~H. (2001).
\newblock Greedy function approximation: a gradient boosting machine.
\newblock {\em Annals of Statistics\/}~(5), 1189--1232.

\bibitem[\protect\citeauthoryear{Friedman and Stuetzle}{Friedman and
  Stuetzle}{1981}]{friedman1981}
Friedman, J.~H. and W.~Stuetzle (1981).
\newblock Projection pursuit regression.
\newblock {\em Journal of the American Statistical Association\/}~{\em
  76\/}(376), 817--823.

\bibitem[\protect\citeauthoryear{Garris and Wilkes}{Garris and
  Wilkes}{2017}]{garris2017}
Garris, M. and B.~Wilkes (2017).
\newblock {Soccernomics: Salaries for World Cup Soccer Athletes}.
\newblock {\em International Journal of the Academic Business World\/}~{\em
  11\/}(2), 103--110.

\bibitem[\protect\citeauthoryear{Groothuis and Hill}{Groothuis and
  Hill}{2013}]{groothuis2013}
Groothuis, P.~A. and J.~R. Hill (2013).
\newblock {Pay discrimination, exit discrimination or both? Another look at an
  old issue using NBA data}.
\newblock {\em Journal of Sports Economics\/}~{\em 14\/}(2), 171--185.

\bibitem[\protect\citeauthoryear{Hamilton}{Hamilton}{1997}]{hamilton1997}
Hamilton, B.~H. (1997).
\newblock Racial discrimination and professional basketball salaries in the
  1990s.
\newblock {\em Applied Economics\/}~{\em 29\/}(3), 287--296.

\bibitem[\protect\citeauthoryear{Hastie, Tibshirani, and Friedman}{Hastie
  et~al.}{2009}]{hastie2009}
Hastie, T., R.~Tibshirani, and J.~Friedman (2009).
\newblock {\em The Elements of Statistical Learning: data mining, inference,
  and prediction}.
\newblock Springer Science \& Business Media.

\bibitem[\protect\citeauthoryear{Hoffer and Freidel}{Hoffer and
  Freidel}{2014}]{hoffer2014}
Hoffer, A.~J. and R.~Freidel (2014).
\newblock {Does salary discrimination persist for foreign athletes in the NBA?}
\newblock {\em Applied Economics Letters\/}~{\em 21\/}(1), 1--5.

\bibitem[\protect\citeauthoryear{Kahn and Shah}{Kahn and Shah}{2005}]{kahn2005}
Kahn, L.~M. and M.~Shah (2005).
\newblock {Race, compensation and contract length in the NBA: 2001--2002}.
\newblock {\em Industrial Relations: A Journal of Economy and Society\/}~{\em
  44\/}(3), 444--462.

\bibitem[\protect\citeauthoryear{Kahn and Sherer}{Kahn and
  Sherer}{1988}]{kahn1988}
Kahn, L.~M. and P.~D. Sherer (1988).
\newblock Racial differences in professional basketball players' compensation.
\newblock {\em Journal of Labor Economics\/}~{\em 6\/}(1), 40--61.

\bibitem[\protect\citeauthoryear{Kelly}{Kelly}{2017}]{kelly2017}
Kelly, T. (2017).
\newblock {Effects of TV Contracts on NBA Salaries}.
\newblock Technical report, Department of Economics, Colgate University, USA.

\bibitem[\protect\citeauthoryear{Lin and Jeon}{Lin and Jeon}{2006}]{lin2006}
Lin, Y. and Y.~Jeon (2006).
\newblock Random forests and adaptive nearest neighbors.
\newblock {\em Journal of the American Statistical Association\/}~{\em
  101\/}(474), 578--590.

\bibitem[\protect\citeauthoryear{Olbrecht}{Olbrecht}{2009}]{olbrecht2009}
Olbrecht, A. (2009).
\newblock Do academically deficient scholarship athletes earn higher wages
  subsequent to graduation?
\newblock {\em Economics of Education Review\/}~{\em 28\/}(5), 611--619.

\bibitem[\protect\citeauthoryear{{R Core Team}}{{R Core Team}}{2020}]{R2020}
{R Core Team} (2020).
\newblock {\em {R: A Language and Environment for Statistical Computing}}.
\newblock Vienna, Austria: R Foundation for Statistical Computing.

\bibitem[\protect\citeauthoryear{Rehnstrom}{Rehnstrom}{2009}]{rehnstrom2009}
Rehnstrom, K. (2009).
\newblock {Racial salary discrimination in the NBA: 2008-2009}.
\newblock {\em Major Themes in Economics\/}~{\em 11\/}(1), 1--16.

\bibitem[\protect\citeauthoryear{Sigler and Compton}{Sigler and
  Compton}{2018}]{sigler2018}
Sigler, K. and W.~Compton (2018).
\newblock {NBA Players' Pay and Performance: What Counts?}
\newblock {\em Sport Journal\/}.

\bibitem[\protect\citeauthoryear{Sigler and Sackley}{Sigler and
  Sackley}{2000}]{sigler2000}
Sigler, K.~J. and W.~H. Sackley (2000).
\newblock {NBA players: are they paid for performance?}
\newblock {\em Managerial Finance\/}~{\em 26\/}(7), 46--51.

\bibitem[\protect\citeauthoryear{Tibshirani}{Tibshirani}{1996}]{tibshirani1996}
Tibshirani, R. (1996).
\newblock {Regression Shrinkage and Selection Via the Lasso}.
\newblock {\em Journal of the Royal Statistical Society: Series B
  (Methodological)\/}~{\em 58\/}(1), 267--288.

\bibitem[\protect\citeauthoryear{Tsagris, Papadovasilakis, Lakiotaki, and
  Tsamardinos}{Tsagris et~al.}{2020}]{tsagris2020}
Tsagris, M., Z.~Papadovasilakis, K.~Lakiotaki, and I.~Tsamardinos (2020).
\newblock {The $\gamma$-OMP algorithm for feature selection with application to
  gene expression data}.
\newblock {\em IEEE/ACM Transactions on Computational Biology and
  Bioinformatics\/}~{\em Accepted for publication}.

\bibitem[\protect\citeauthoryear{Vincent and Eastman}{Vincent and
  Eastman}{2009}]{vincent2009}
Vincent, C. and B.~Eastman (2009).
\newblock {Determinants of pay in the NHL: A quantile regression approach}.
\newblock {\em Journal of Sports Economics\/}~{\em 10\/}(3), 256--277.

\bibitem[\protect\citeauthoryear{Wen}{Wen}{2018}]{wen2018}
Wen, R. (2018).
\newblock {Does Racial Discrimination Exist Within the NBA? An Analysis Based
  on Salary-per-Contribution}.
\newblock {\em Social Science Quarterly\/}~{\em 99\/}(3), 933--944.

\bibitem[\protect\citeauthoryear{Wiseman and Chatterjee}{Wiseman and
  Chatterjee}{2010}]{wiseman2010}
Wiseman, F. and S.~Chatterjee (2010).
\newblock Negotiating salaries through quantile regression.
\newblock {\em Journal of Quantitative Analysis in Sports\/}~{\em 6\/}(1).

\bibitem[\protect\citeauthoryear{Wright and Ziegler}{Wright and
  Ziegler}{2017}]{ranger2017}
Wright, M.~N. and A.~Ziegler (2017).
\newblock {ranger: A Fast Implementation of Random Forests for High Dimensional
  Data in C++ and R}.
\newblock {\em Journal of Statistical Software\/}~{\em 77\/}(1), 1--17.

\bibitem[\protect\citeauthoryear{Xiong, Greene, Tanielian, and Ulibarri}{Xiong
  et~al.}{2017}]{xiong2017}
Xiong, R., M.~Greene, V.~Tanielian, and J.~Ulibarri (2017).
\newblock Research on the relationship between salary and performance of
  professional basketball team (nba).
\newblock In {\em Proceedings of the 8th International Conference on
  E-business, Management and Economics}, pp.\  55--61.

\bibitem[\protect\citeauthoryear{Yang and Lin}{Yang and Lin}{2012}]{yang2012}
Yang, C.-H. and H.-Y. Lin (2012).
\newblock {Is there salary discrimination by nationality in the NBA? Foreign
  talent or foreign market}.
\newblock {\em Journal of Sports Economics\/}~{\em 13\/}(1), 53--75.

\bibitem[\protect\citeauthoryear{Yilmaz and Chatterjee}{Yilmaz and
  Chatterjee}{2003}]{yilmaz2003}
Yilmaz, M. and S.~Chatterjee (2003).
\newblock {Salaries, performance, and owners' goals in major league baseball: A
  view through data}.
\newblock {\em Journal of Managerial Issues\/}~{\em 15\/}(2), 243--255.

\bibitem[\protect\citeauthoryear{Zimmer and Zimmer}{Zimmer and
  Zimmer}{2001}]{zimmer2001}
Zimmer, M.~H. and M.~Zimmer (2001).
\newblock {Athletes as Entertainers: A Comparative Study of Earnings Profiles}.
\newblock {\em Journal of Sport and Social Issues\/}~{\em 25\/}(2), 202--215.

\end{thebibliography}
\end{document}